\documentclass[1p, 12pt, preprint, times]{elsarticle}

\usepackage{lineno}%,hyperref}
\journal{Journal of \LaTeX\ Templates}

\usepackage[applemac]{inputenc}
\usepackage{xspace}
\usepackage[inline]{enumitem}
\usepackage[version=4]{mhchem} % \ce{ }
\usepackage{threeparttable}
\usepackage{amsmath,amssymb}%      pour les maths
\usepackage{adjustbox}
\usepackage[dvipsnames]{xcolor} % pour  les coculeurs
\newcommand{\via}{\textit{via}\xspace}
\newcommand{\vs}{\textit{vs}\xspace}
\newcommand{\etal}{\textit{et al.}\xspace}
\newcommand{\tiod}{\ce{TiO2}\xspace}
\newcommand{\srtio}{\ce{SrTiO3}\xspace}
\newcommand{\titp} {\ce{Ti^3+}\xspace}
\newcommand{\ppod}{p(\ce{O2})\xspace}
\newcommand{\mum}{\;$\mu$m\xspace}
\newcommand{\mm}{\;mm\xspace}
\newcommand{\thz}{\;THz\xspace}
\newcommand{\ghz}{\;GHz\xspace}
\newcommand{\neff}{n_{eff}}
\newcommand{\neffmin}{n_{eff_{min}}\xspace}
\newcommand{\phonon}{\omega_{T_{O_1}}\xspace}
\newcommand{\gamphonon}{\gamma_{T_{O_1}}\xspace}
\newcommand{\tandel}{\tan\delta}
\newcommand{\epsp}{\varepsilon'}
\newcommand{\epspp}{\varepsilon''}
\newcommand{\epseff}{\varepsilon_{eff}}
\newcommand{\epsr}{\varepsilon_r}
\newcommand{\epspar}{\varepsilon_\parallel}
\newcommand{\epsperp}{\varepsilon_\perp}
\newcommand{\mueff}{\mu_{eff}}

\newcommand{\Deltafs}{\Delta f_s}
\newcommand{\Deltafc}{\Delta f_{Cohn}}
\newcommand{\txcb}[1]{\textcolor{RoyalBlue}{#1}}

\newcommand{\dix}[1]{\cdot10^{#1}}
\newcommand{\cels}[1]{#1\,\textcelsius\xspace}
\newcommand{\fig}[1]{Fig.\,\ref{#1}}
\newcommand{\tab}[1]{Table\,\ref{#1}}
\newcommand{\seef}[1]{(see~Fig.\,\ref{#1})}
\newcommand{\seeref}[1]{(see ref.\,\cite{#1})}
\newcommand{\seetab}[1]{(see Table\,\ref{#1})}
\newcommand{\seefsm}[1]{(see~Fig.\,\ref{#1} in the supplementary material)}
\newcommand{\seetabsm}[1]{(see~table.\,\ref{#1} in the supplementary material)}
\newcommand{\seefsmd}[2]{(see~Fig.\,\ref{#1}\,\&\,\ref{#2} in the supplementary material)}% in the supplementary material)}

\newcommand{\seeeqd}[2]{(see eq.\,\eqref{#1}\,\&\,\eqref{#2})}

\newcommand{\dbet}{$D_{BET}$\xspace}
\newcommand{\sbet}{$S_{BET}$\xspace}

\newcommand{\ts}{T$_{SPS}$}
\newcommand{\tsps}{T$_{SPS}$}
\newcommand{\tann}{T$_{ann}$}
\newcommand{\tcs}{T$_{CS}$}
\newcommand{\tuu}{$TE_{11}$\xspace}
\newcommand{\tud}{$TE_{12}$\xspace}
\newcommand{\fuu}{$f_{11_{s}}$\xspace}
\newcommand{\fud}{$f_{12_{s}}$\xspace}
 \newcommand{\ra}[1]{\renewcommand{\arraystretch}{#1}} % linespacing in booktable
\usepackage{xr}
\usepackage{xr-hyper}
\usepackage{hyperref}
\usepackage{booktabs}
\usepackage{tablefootnote}
\usepackage{subcaption}
\usepackage{float}
\usepackage[T1]{fontenc} % ---> les accents ? 
\usepackage{amsmath}
\usepackage{microtype}
\usepackage{xr}
\usepackage{xr-hyper}
\usepackage{hyperref}
\date{\today}
\selectfont
\begin{document}

\begin{frontmatter}

\title{Impact of sintering conditions on the dielectric properties of \tiod ceramics for metamaterials applications at terahertz frequencies.}
%\title{Elsevier \LaTeX\ template\tnoteref{mytitlenote}}

%\author[label1]{Djihad Amina Djemmah} %% Author name
%\author[label2]{Delphine Gourdonnaud}
%\author[label1]{Ludovic Lazrgeau}
%\author[label3]{Fayçal Bouamrane}
%\author[label4]{Jean-François Roux}
%\author[label2]{Pierre-Marie Geffroy}
%\author[label1]{Eric Akmansoy\corref{mycorrespondingauthor}}
\author{Djihad Amina Djemmah\fnref{label1}}
\author{Delphine Gourdonnaud\fnref{label2}}
%\author{Ludovic Largeau\fnref{label1}}
\author{Fayçal Bouamrane\fnref{label3}}
\author{Jean-François Roux\fnref{label1}}
\author{Pierre-Marie Geffroy\fnref{label2}}
\author{\'Eric Akmansoy\corref{cor1}\fnref{label1}}
\cortext[cor1]{eric.akmansoy@universite-paris-saclay.fr}
\affiliation[label1]{
organization={Centre de Nanosciences et Nanotechnologies - CNRS (UMR 9001) - Université Paris-Saclay}, %(C2N) - UMR 9001 - Universit\'e Paris-Saclay},%Department and Organization
            addressline={10 bd Thomas Gobert}, 
            city={Palaiseau},
            postcode={91120}, 
            country={France}
            }
 \affiliation[label2]{
 organization={Centre Européen de la Céramique - CNRS (UMR 7315) - Université de Limoges},
             addressline={12 Rue Atlantis},
             city={Limoges},
             postcode={87068 LIMOGES Cedex Atlantis},
             country={France}
             }
 \affiliation[label3]{
organization={Laboratoire Albert Fert - CNRS, (UMR137) - Thales, Université Paris-Saclay}, %- CNRS, Thales, Université Paris-Saclay},
             addressline={1 Avenue Augustin Fresnel},
             city={Palaiseau},
             postcode={91120},
             country={France}
}

 \affiliation[label4]{
 organization={Centre de Radiofrequences, Optique et Micro-nanoélectronique des Alpes - CNRS (UMR 5130 ) - Université Savoie Mont Blanc},
             addressline={rue Lac de la Thuile, Bat. 21},
             city={Le Bourget du Lac},
             postcode={73370},
             country={France}
             }

%\tableofcontents
%\newpage
%\listoffigures
%\newpage

\begin{abstract}
Titanium dioxyde (\tiod) is a promising dielectric material for the realization of metamaterials operating in the terahertz (THz) range. 
Indeed, these necessitate a high permittivity and low loss material. 
%These necessitate to structure the bulk material at the tens of micrometers scale with a micrometer resolution. 
In this paper, we compare the processes of fabrication and the results of characterisation of bulk \tiod pellets. From the results of this characterization, we have numerically designed 2D all dielectric metamaterials (ADM) showing that they may exhibit negative or near-zero effective index. Our previous simulations show that the relative permittivity $\epsp$ has to be around 100, while the loss tangent $\tandel$ has to be lower than $0.02$. We have thus compared conventional sintering (CS) \vs spark plasma sintering (SPS), and investigated the effect of post-sintering annealing on the loss to satisfy these two criteria. The samples were characterized by THz Time Domain Spectroscopy (THz-TDS). One of the samples exhibits a loss tangent as low as $\tandel = 6\cdot 10^{-3}$, with a permittivity $\epsp=103$. These results highlight the importance of the fabrication process on the EM properties of bulk \tiod, and demonstrate that it is a promising material for the development of metamaterial in the\thz band.
\end{abstract}

\begin{keyword}
Ceramics processes, \tiod, THz-Time Domain Spectroscopy, All-Dielectric Metamaterials.
%\texttt{elsarticle.cls}\sep \LaTeX\sep Elsevier \sep template
%\MSC[2010] 00-01\sep  99-00
\end{keyword}

\end{frontmatter}

%\linenumbers

% -------------------------------------------------------
% --- Introduction
% -------------------------------------------------------
\section{Introduction}
Ceramics are of great interest as the realization of metamaterials (MM) operating in the THz range is aimed.
All-Dielectric Metamaterial (ADM) are perfectly suited to this frequency range\;\cite{im1_luo, ttst3_takano}, unlike traditional MMs whose unit cell is metallic and suffer from increasing ohmic loss with the frequency. Moreover, the unit cell of ADMs, which consists of two High Permittivity Resonators (HPR), is of simple geometry. As the incident EM field illuminates the ADM, the first two Mie resonances are excited, which may lead to negative or near-zero effective index\;\cite{Veselago:2003, opn27_engheta}. To sustain these resonances, the material of the HPR must have a high permittivity and low loss, their dimensions being sub-wavelength. \tiod and \srtio are good candidates to satisfy these two conditions. \tiod provides a good compromise between high permittivity and low loss, notably because if its permittivity is lower than that of \srtio,  its first optical phonon $\phonon$ is at 5.7\thz, whereas that of \srtio is at 2.7\thz\cite{pr126_spitzer, pr145_barker, sr8_dupas}. This means that the loss of \tiod are lower in the considered frequency range. About \tiod, our previous simulations show that to get the first two Mie resonances, two criteria --- high permittivity ($\epsr \simeq 100$) and low loss ($\tandel \leq 0.02$) ---  have to be satisfied\;\cite{djemmah2022processing, ijmwt_amina}.
 
We therefore fabricated bulk \tiod pellets by two different processes: conventional sintering (CS) and spark plasma sintering (SPS), and compared their dielectric properties. We also investigated the effect of post-sintering annealing, showing its influence on the EM properties of the material. We characterized the pellets by the means of THz Time Domain Spectroscopy (THz-TDS), and then numerically designed 2D ADMs from these experimental results: negative or near-zero effective index is achievable.

\section{Methods}
% -------------------------------------------------------
%--- Fabrication of the ceramic
% -------------------------------------------------------
\subsection{Fabrication of the ceramics}
% -------------------------------------------------------
% Starting  \tiod powders
% -------------------------------------------------------
\subsubsection{Starting \tiod powders}
Fifty five different samples have been fabricated to address the dependance of the EM properties (permittivity and loss)  on several parameters: density, type of  sintering, sintering and annealing temperatures. These sintering studies of \tiod were conducted using commercial powders, whose particle size \dbet was determined from the specific surface \sbet with the relative density $\rho = 4.23\;g\cdot cm^{-3}$ of \tiod \via the Brunauer-Emmett-Teller' method (BET)\;\cite{pst32_courard}.  %\SI{6.022e23}{\per\mol}
Before sintering, the commercial powders need to be compacted into green parts by uniaxial pressing. %The \tiod powder is thus loaded into a rigid die and compacted by applying a uniaxial pressure (50 MPa).

% -------------------------------------------------------
% ------------Elaboration and sintering of the \tiod samples
%-------------------------------------------------------
\subsubsection{Elaboration and sintering of the \tiod samples}
Bulk \tiod ceramic was shaped by two different sintering processes from the green parts: conventional sintering (CS) and Spark Plasma Sintering (SPS). Eighteen samples were conventionally sintered, while thirty seven were sintered by SPS.
During the former process, the \tiod powder is first compacted in a 10\mm diameter die under a 50 MPa uniaxial pressure to produce green compacts. These have been then sintered at various high temperatures: \tcs=\cels{1300}, \cels{1350},  \cels{1400},  \cels{1450}, \cels{1500} and \cels{1550} for about two hours, {under argon atmosphere corresponding to the oxygen partial pressure $\ppod = 2.1\dix{-4}$\;Pa. }% for 30 min to obtain the dense material.
%In the conventional sintering process, \tiod powder is compacted using a die of 10\mm diameter under a compressive pressure of 50 MPa to form green compacts. These compacts are then sintered at 1550°C during 30 min to produce a dense material.
%
%During the latter process, the sintering of the \tiod powder, placed in a 8\mm graphite die, is rapidly performed by a high intensity pulsed electric current (about 10\;kA). A 10\;MPa pressure is applied along the vertical axis of the sample by two pistons for about ten minute in a vacuum chamber???. The atmosphere can be neutral (Ar), reducing (N2) or a primary vacuum (a few Pa).  (What's our case???) 
During the latter process, the sintering of the \tiod powder, placed in 8\mm graphite die, is rapidly performed by a high intensity pulsed electric current (about 10\;kA). A 10\;MPa pressure is applied along the vertical axis of the sample by two pistons during 5 min in a vacuum chamber. The atmosphere is neutral, namely argon at a pression of a few Pascal.  
The temperature is controlled \via an IR pyrometer. SPS allows to sinter the powder at lower temperatures than conventional sintering and is a shorter process. It has been carried out at various temperatures: \ts=\cels{1100}, \cels{1150},  \cels{1200} and \cels{1300}. A Dr. Sinter 825 SPS machine (by Fuji Electronics Industrial Co. Ltd) was used. 
%The sintering process is shorter that the conventional one and its operating temperature is lower.

After the sintering, some of the samples have been annealed, at various temperatures (\tann=\cels{950}, \cels{1000}, \cels{1100}, \cels{1200}, \cels{1300} and \cels{1550}), {under air atmosphere also corresponding to the oxygen partial pressure $\ppod = 2.1\dix{-4}$\;Pa.}  %between \cels{950} and \cels{1100} for 60 h???. 
Next, the \tiod pellets were thinned by polishing up to thicknesses from 300 to 500\mum. Photographs of the pellets are reported in the supplementary material \seefsm{S-fig:photo}

% -------------------------------------------------------
% Characterization
% -------------------------------------------------------
\subsection{Characterization}
The structural properties of the fabricated samples were investigated by Scanning Electron Microscope (SEM) and X-ray diffraction (XRD). Next, the samples were characterized at THz frequencies by the means of Time Domain Spectroscopy (THz-TDS).
% -------------------------------------------------------
% XRD
% -------------------------------------------------------
\subsubsection{XRD}
Titanium dioxide (\tiod) crystallizes in three main forms, two of which may have EM applications: rutile and anatase. These have different crystalline structures, which results in different electronic and optical properties. Because of its higher permittivity and lower loss, the rutile structure is better suited for metamaterials applications. To determine which structure has resulted from our fabrication processes, we have characterized the samples by X-ray diffraction (XRD). The 2Theta/Theta (2$\Theta$/$\Theta$) scans were collected in a Bragg-Brentano geometry using a Rigaku Smartlab diffractometer.% Two results are reported in  \fig{fig:xrd}: that of a conventionally sintered sample and the other of a SPS one. The former has been sintered at \cels{1350} while the latter has been at \cels{1100}. The 2Theta/Theta scans were collected in a Bragg-Brentano geometry using a Rigaku Smartlab diffractometer. 

\subsubsection{Density}
The density of the sintered samples was measured using Archimede's method\;\cite{jjap33_sasaki}.  %while the relative permittivity was determined through dielectric characterization at room temperature
% -------------------------------------------------------
% --- THz caracterisations
% -------------------------------------------------------
\subsubsection{THz characterisations}
Once the pellets fabricated and polished, they were characterized in the [0.2, 1.8] THz range by the means of THz Time Domain Spectroscopy (THz-TDS). {Actually, this broadband technique allows to determine the dielectric permittivity (including both real and imaginary parts) of the material if this latter is sufficiently transparent. Moreover, in the case of inhomogeneous media or compounds one can also have an insight about structural parameters such as porosity using an effective medium analysis \seeref{jecs42_hakobyan} %(Je l’ai rajoutée pour avoir une reference qui cite JECS, il faut la numérotée mettre comme il faut). 
In our case we performed transmission experiment using a commercial Terapulse Lx spectrometer (from Teraview\texttrademark). }
In this system, the sample is placed at the focal point of two parabolic mirrors where the frequency dependent waist of the THz beam decreases from 2\mm at 0.2 THz to 0.2\mm at 2 THz. The spectrometer delivers THz pulses whose spectrum covers the [0.2, 5]\thz band \seefsmd{S-fig:tds_td}{S-fig:spectre}. The dynamic range of measurement tops around 0.6\thz and is gradually attenuated at high frequencies; the signal reaches the noise level around 5\thz. Towards the lower end of the spectrum, the amplitude of the detected signal is mainly limited by diffraction effects as the low-frequency components of the\thz beam are poorly focused towards the detector and may even be blocked by the sample holder if the size of sample is too small. Considering these system parameters, we have fabricated pellets with diameter ranging from 8 to 13\mm. As mentioned above, before being characterized, each sample is thinned into a parallel-sided plate with a thickness between 300 and 500\mum. Given the expected absorption losses of the samples, which generally range between $\tandel= 0.1$ and $\tandel$= 0.2 to at 1.8\thz, such thicknesses would induce propagation losses of approximately 40 to 65 dB. As a consequence, the upper limit of the frequency range of characterization of the ceramic pellets varies from 1 to 1.8\thz depending on the exact parameters of the sample. Generally, the lower limit of reliable characterization is in between 0.2 and 0.25\thz. In order to reduce the impact of\thz water absorption onto the experiment, the samples are placed in a chamber purged by nitrogen gas with a residual humidity rate below 10\% at ambient temperature. Each measurement consists of recording time traces with a time equivalent length of 30 ps. {For further numerical analysis of the signal in the frequency domain, this window width will lead to a frequency resolution of 33 GHz which is good enough to characterize material, such as \tiod pellets, that does not present sharp resonances in this frequency range. We note that the characterization of resonant devices such as metamaterial with narrow resonances would require longer acquisition window or the use of a frequency resolved\thz apparatus.}

{In practical terms}, to characterize the pellets, we follow a classical method based on the comparison of signals recorded with or without the sample\;\cite{roux2014principles}. Firstly, a reference temporal signal without the sample is recorded; secondly, the signal transmitted by the sample is recorded. The two measured temporal signals are then processed by a Fast Fourrier Transform (FFT), which yields the transmission spectrum. {The time domain signals and the spectra are reported in the supplementary material \seefsmd{S-fig:tds_td}{S-fig:spectre}.} The calculation leads to the knowledge of the complex transmission coefficient of the sample. From this and the sample thickness, the complex refractive index $n +  \imath\kappa$ is extracted assuming that the signal does not suffer from any scattering or diffraction effects. From the refractive index $n$ and the extinction coefficient $\kappa$, the absorption coefficient $\alpha$ is deduced, from which, in turn, the complex dielectric function $\varepsilon^* = \epsp + \imath\epspp$ is found. All this analysis is done using a dedicated numerical tool developed using Matlab\texttrademark\; and based on the method originally described by Duvillaret \etal\;\cite{duvillaret1996reliable}.

\subsection{Design of the All Dielectric Metamaterials and numerical simulations}\label{dadm}
% -------------------------------------------------------
% --- Unit Cell ADM sketch
% -------------------------------------------------------

\begin{figure}[ht]
\begin{center}
\includegraphics[width = \textwidth]{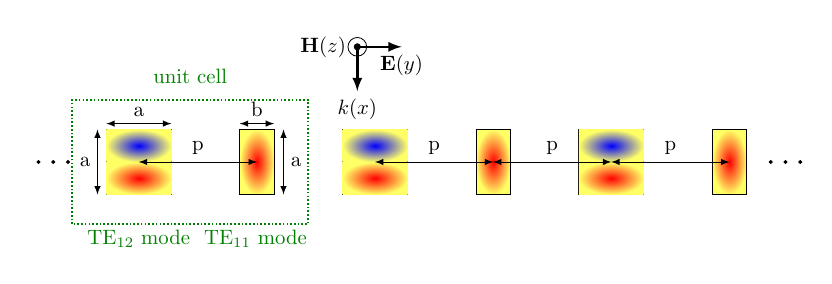}
\caption{Sketch of a 2D All Dielectric Metamaterial (ADM). It is a periodical structure, whose unit cell consists of two High Permittivity Resonators sustaining the first two modes of Mie resonances, as the incident field is Transverse Electric (the field $\vec{E}$ is perpendicular to the axis of the resonators). Their cross-sections are $a\times a$ and $a\times b$, while the half-period $p$ is the distance between the axes of two resonators. Both the resonators are infinite along the $z$ direction. The two modes are shown. The \tuu mode exhibits one antinode while the \tud has two\;\cite{jap109_lepetit}.
}
\label{fig:unit_cell}
\end{center}
\end{figure}
% -------------------------------------------------------

{Given that we intend to develop} ADMs operating in the\thz from the ceramics we fabricated, we used the results of the\thz caracterization of our pellets to design and numerically simulate ADMs. These are periodic structures whose unit cell consists of two High  Permittivity Resonators, sustaining the first two modes of the Mie 	resonances. 
The small resonator sustains the first mode of the Mie resonances, while the other sustains the second mode. The former is called the ``magnetic resonator'' and the latter the ``electric resonator''.
The first one gives rise to resonant effective permeability $\mueff$, while the second one gives rise to resonant effective permittivity $\epseff$, both of which may take negative values. Their superposition may then result in a negative effective index or near zero-index\;\cite{jap109_lepetit, apl95_lepetit}. Fig.\;\ref{fig:unit_cell} reports the sketch of a 2D ADM. 

We designed 2D ADMs operating at 0.3\thz and 0.6\thz, because this corresponds to transmission windows of the atmosphere. The dimensions of the sides of the two rectangular resonators set their frequency of resonance, according to the Cohn's model\;\cite{jap109_lepetit, apl95_lepetit}, which provides the frequency of the modes by the relation\;\cite{mtt16_cohn}: 
\begin{equation}\label{eq:cohn}
f_{mn} = \frac{c}{2\sqrt{\epsp}} \sqrt{\left(\frac{m}{a } \right)^2 + \left(\frac{n}{b} \right)^2}, 
\end{equation}
where $m$ and $n$ are integers, and {$a$} and {$b$} are the dimensions of the sides of the 2D rectangular resonators. It consequently depends on the relative permittivity $\epsp$ of the bulk material. Since we consider the first two modes of Mie resonances, $m = n = 1$ and $m=1$, $n=2$ in the elementary cell, the first mode (denoted \tuu) being the magnetic one and the second (denoted \tud), the electric one. Notice that the Cohn's model provides the frequencies of resonances of an isolated resonator. 

The dimensions of the resonators are therefore adjusted so that the two modes overlap, resulting in the superposition of effective permeability $\mueff$ and effective permittivity $\epseff$. Operating at 0.3\thz, the magnetic resonator has a rectangular cross section (a=110\mum$\times$ b=58\mum), while the electric one has a square cross section\;(a=110\mum$\times$ a=110\mum). To excite the two modes, the polarization of the incident EM field is Transverse Electric (TE), that is, the electric field $\mathbf{E}$ is perpendicular to the axis of the dielectric resonators, while the magnetic one $\mathbf{H}$ is parallel to the axis. 
From Cohn's model, the respective resonance frequencies are 0.287\;THz (\tuu mode) and 0.300\;THz (\tud mode).
%The respective resonance frequencies are 0.282\;THz (\tuu mode) and 0.294\;THz (\tud mode).
The two modes undergo  coupling,  which notably moves the frequencies of resonances that are given by the Cohn's model, and can lead to the degenerescence of the modes, that is, their frequencies become equal as $p$ decreases\;\cite{epjam5_marcellin}. The period of the lattice $2p$ plays thus an important role in the EM characteristics; it is set as $p=200$\mum. 
Operating at 0.6\thz, %both the resonators have a rectangular cross section and their 
the sides of the resonators are respectively b=28\mum$\times$ a=55\mum and a=55\mum$\times$ a=55\mum, while the lattice period is set as $p=120$\mum. 
Their respective resonance frequencies are 0.586\;THz (\tuu mode) and 0.594\;THz (\tud mode).
Therefore, the cross-sections of the resonators forming an ADM are sub-wavelength. Varying the lattice period $2p$ changes the effective index $\neff$, so that near-zero index ($0\leq\neff\leq1$) is achievable\;\cite{ijmwt_amina, epjam5_marcellin}.
%The frequencies of resonance of course depends on the  relative permittivity of the bulk material.??? 
%
 %Our previous simulations demonstrate that to get a negative index requires the loss tangent $\tandel$ being lower than 0.02\;\cite{djemmah2022processing}.  
Our previous simulations show that the relative permittivity has to be around  $\epsp \simeq 100$, while the loss tangent $\tandel= \epsilon'/\epsilon''$ has to be lower than 0.02, so as to obtain the two resonances\;\cite{ijmwt_amina, djemmah2022processing}. % \textbf{Section 4.4}??? 
%To design the device, \textbf{we used the experimental EM characteristics of sample 27}, that was sintered by SPS at \cels{1150} and annealed at \cels{1000}under air during sixty hours. These are ($\epsp = 107, \tandel = 0.006$) at 0.3\thz and  ($\epsp = 108, \tandel = 0.016$) at 0.6\thz, respectively.
%
The simulations have been carried out by the means of the Lumerical\texttrademark\;commercial software, which involves the Finite-difference time-domain (FDTD) method and yields the S-parameters (reflexion and transmission coefficients) \seefsm{S-fig:param_s}, from which we extracted the effective parameters: permeability $\mueff$, permittivity $\epseff$ and index $\neff$\;\cite{ jap109_lepetit, apl95_lepetit}, using Szabo \etal method\;\cite{mtt58_szabo}. 

% -------------------------------------------------------
% --- result and discussion
% -------------------------------------------------------
\section{Results and discussion}
\subsection{Chemical properties of the \tiod ceramics.}
% ------------------------------------------------
% --- SEM images
% ------------------------------------------------
\subsubsection{SEM images}
Scanning Electron Microscope (SEM) images of the polycristaline pellets were taken, from which the grain size and the morphology of the sintered ceramic can be observed\:\seef{fig:sem}: 
%
%and some images are shown in \fig{fig:sem}, these show the size of the  
%from which it can be observed that t
the SPS \tiod samples have a smaller grain size (4-6\mum) with very low residual porosity (close to 1.2\%), than that of the CS samples that have a larger grain size (10-30\mum), with intragranular and intergranular porosity, that is close to 3.8\%. That said, the grain size has no significant effect on the permittivity and the dielectric loss, specially when the ceramic powders are very pure\;\cite{jecs36_geffroy}. The determination of the porosity ratio and the mean grain size have been performed from the SEM images by ImageJ software.

%while the \tiod samples sintered by conventional sintering (CS) at \cels{1550} show large grains size (20-50\mum), with intragranular and intergranular porosity, that is close to 2-3\%.
%({Pierre-Marie et Delphine, it's up to you !})

\begin{figure}[ht]
\centering
\begin{subfigure}{0.6\textwidth}
    \includegraphics[width=\textwidth]{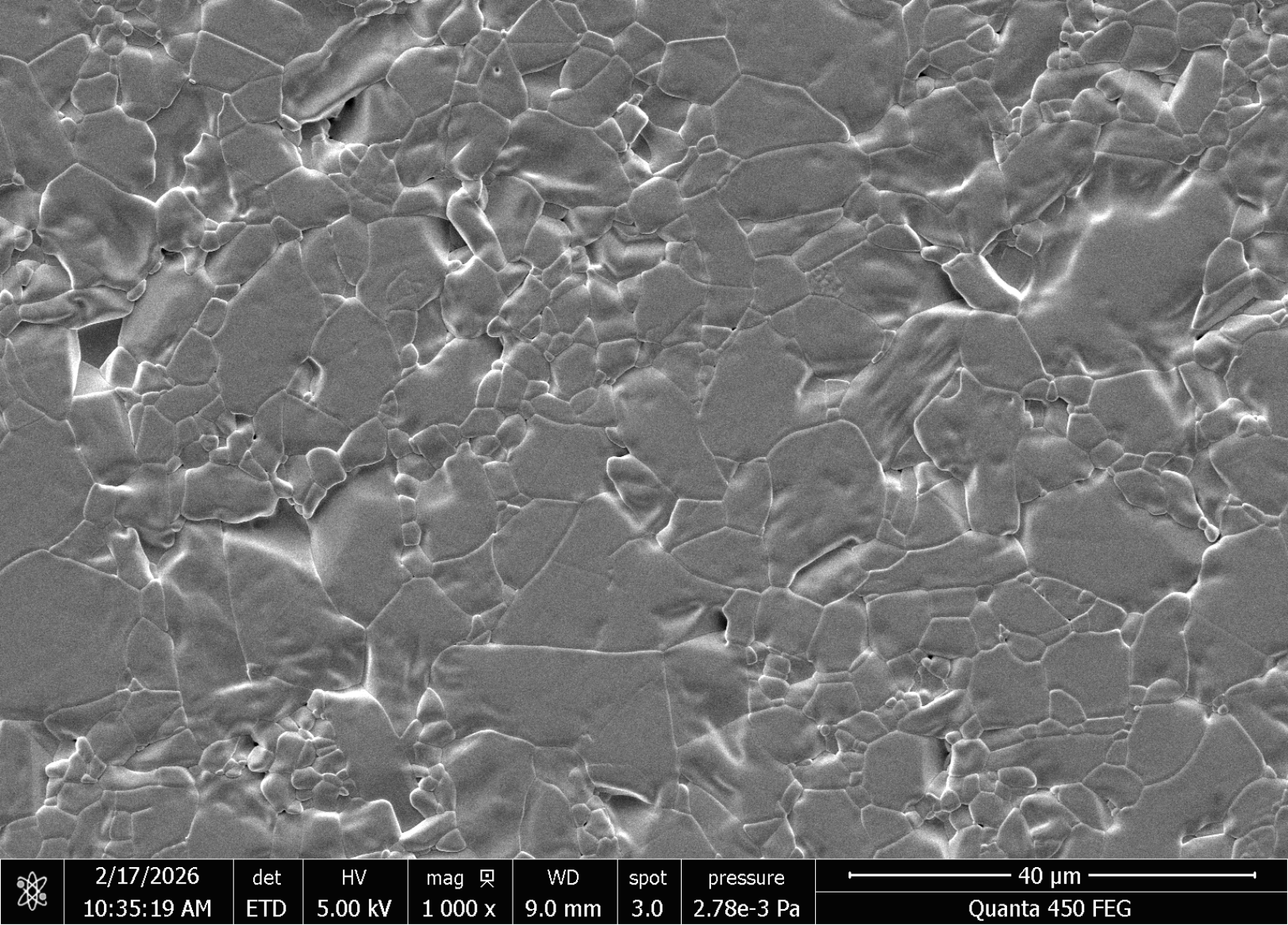}

    \caption{}
    \label{fig:sem_sps}
\end{subfigure}
\\
\begin{subfigure}{0.6\textwidth}
    \includegraphics[width=\textwidth]{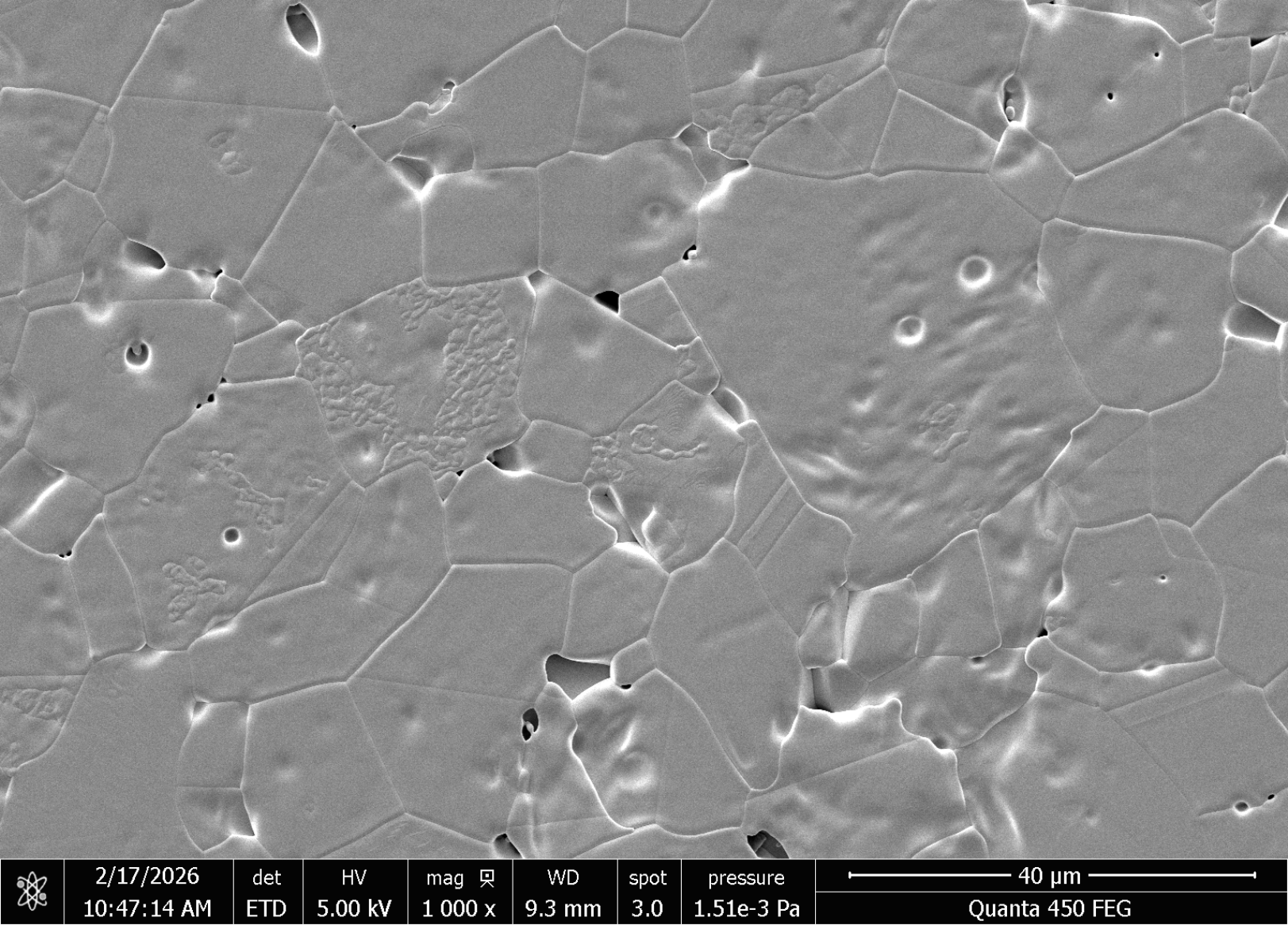}
    \caption{}
    \label{fig:sem_cs}
\end{subfigure}
\caption{SEM microphotographies of \tiod samples sintered (a) by spark plasma sintering (SPS) at \cels{1100} 
and (b) by conventional sintering at \cels{1550}.
}
%\caption{SEM images of sintered \tiod samples showing their microstructures: (a) Sample sintered by SPS (\cels{1100} during 10 min). (b) Conventionally sintered sample (\cels{1550} during 2h).}
\label{fig:sem}
\end{figure}

% ------------------------------------------------
% --- XRD
% ------------------------------------------------
\subsubsection{XRD results}
\begin{figure}[ht]
\centering
\begin{subfigure}{0.75\textwidth}
    \includegraphics[width=\textwidth]{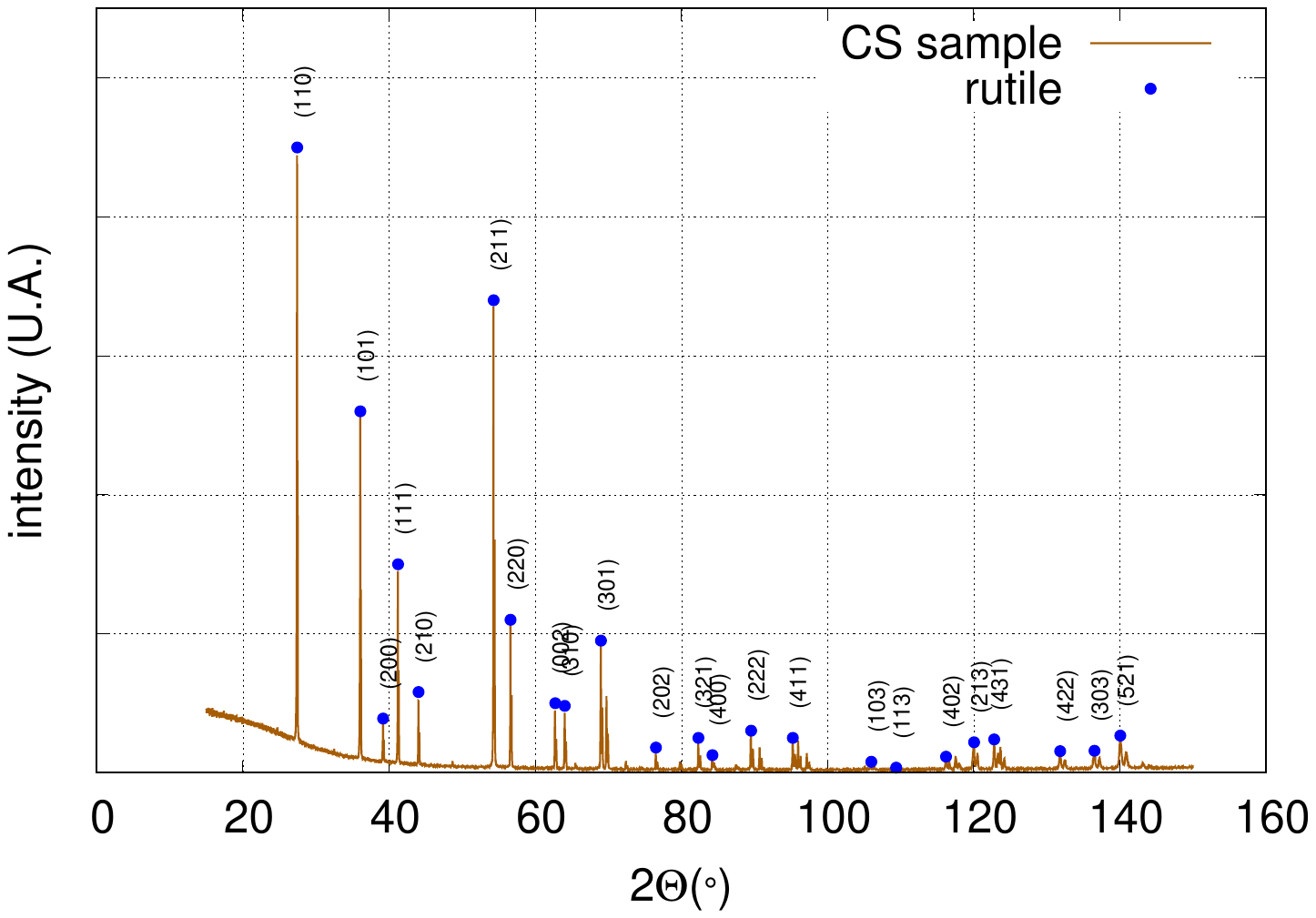}
    \caption{}
    \label{fig:xrd12}
\end{subfigure}
\\
\begin{subfigure}{0.75\textwidth}
    \includegraphics[width=\textwidth]{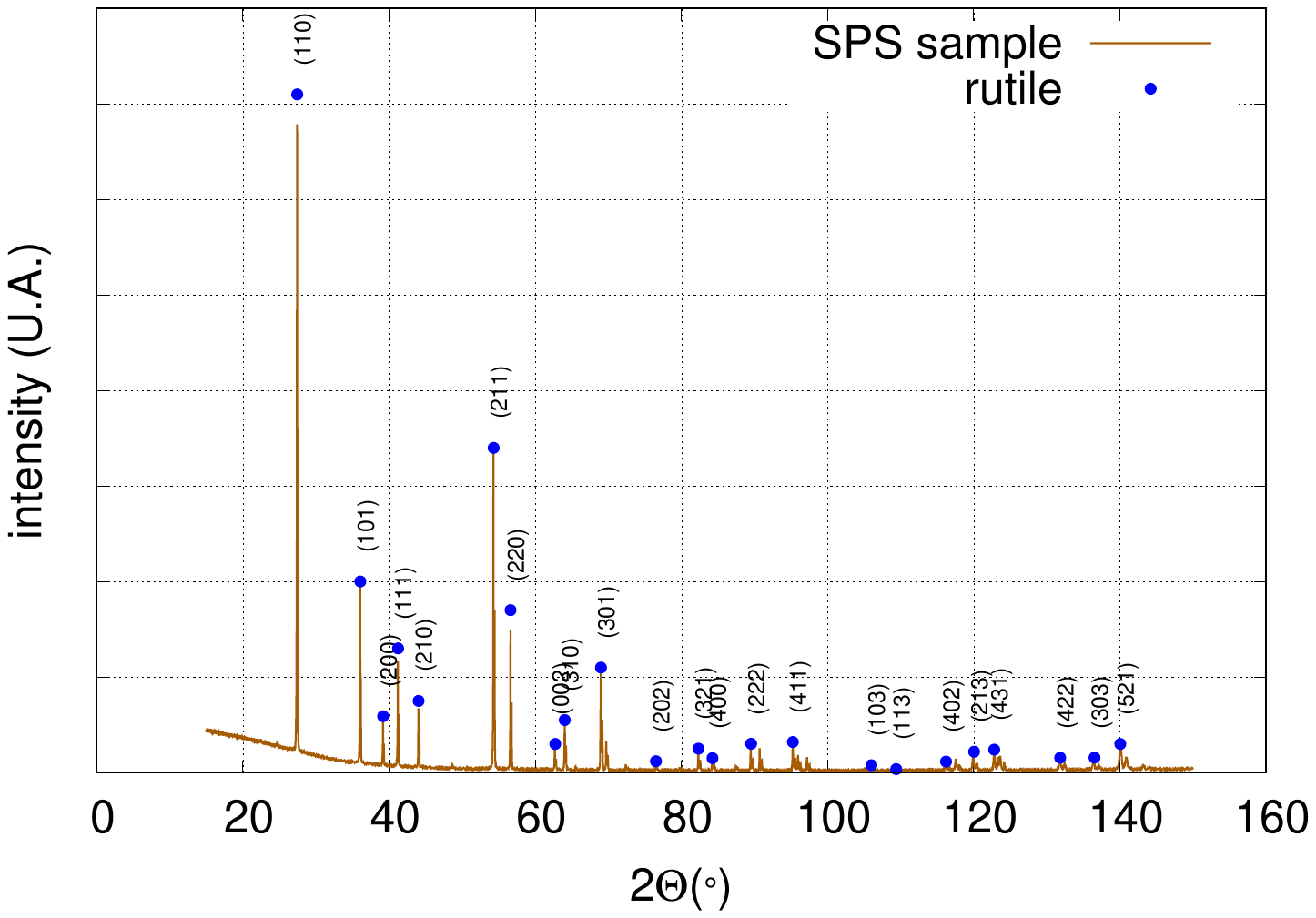}
    \caption{}
    \label{fig:xrd31}
\end{subfigure}
\caption{XRD spectra of two \tiod ceramics samples: (a) Conventionally sintered at \tcs=\cels{1350} (CS sample), (b) Sintered by SPS at \tsps=\cels{1100} (SPS sample) compared to the rutile spectrum: the Miller indexes of the \tiod rutile form are the blue marks\;\cite{icdd}. }
\label{fig:xrd}
\end{figure}

Two results of XRD characterizaiton are reported in \fig{fig:xrd}: that of a CS sample and the other of a SPS one. The former has been sintered at \tcs=\cels{1350}, while the latter has been at \tsps=\cels{1100}. 
These  spectra are compared with that of rutile \tiod, which confirms that the samples are of the rutile form\;\cite{icdd}.

% -------------------------------------------------------
% --- Dielectric Properties of TiO2 Ceramic
% -------------------------------------------------------
\subsection{Dielectric Properties of the \tiod Ceramics}
We addressed the EM properties of the pellets in function of the density, the type of  sintering, the sintering and annealing temperatures, and then the dielectric function of the \tiod ceramics.
\subsubsection{Permittivity \vs density}
\begin{figure}[ht]
\begin{center}
\includegraphics[width = 0.8\textwidth]{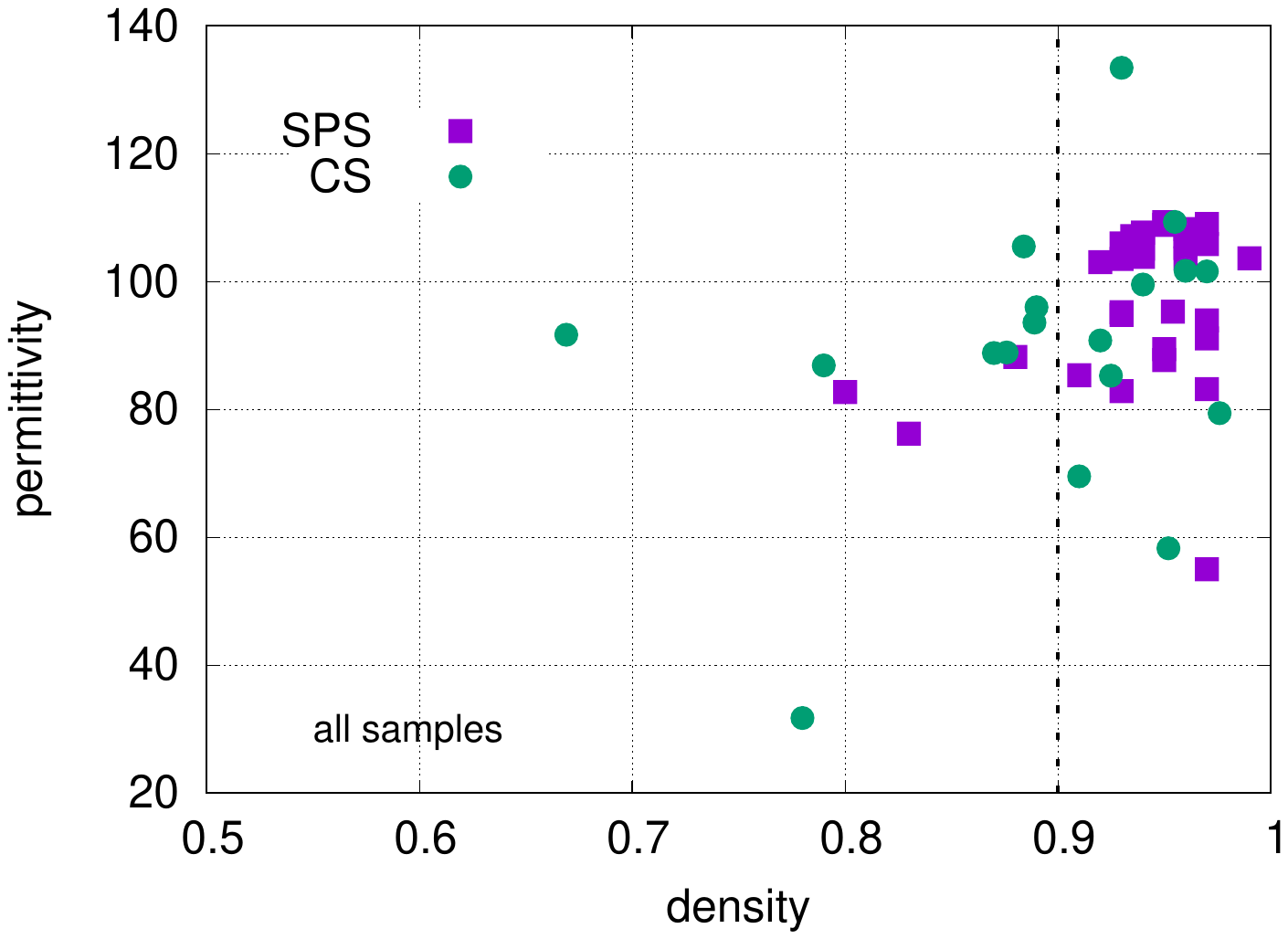}
\caption{Permittivity $\epsp$ of all the samples in function of their density: SPS samples in violet and CS samples in green. The vertical dotted line indicates the density of 0.9.  }
\label{fig:permittivity_density}
\end{center}
\end{figure}

The permittivity of all the samples \vs their relative density is reported in \fig{fig:permittivity_density}, and it can be seen that both the processes allow to fabricate dense samples (density > 0.9) with high permittivity ($\varepsilon' \simeq 100$). And it can be noticed that most of high density samples have a high permittivity, as might be expected\;\cite{sr8_dupas}. 
Nonetheless, most of the SPS samples have a density greater than 0.9 and a permittivity around $\epsr \simeq$100 . %, which confirms our previous results\;\cite{sr8_dupas}. 

% ------------------------------------------------
% ----  permittivity of all samples
% ------------------------------------------------

%
% ------------------------------------------------
% --- loss tangent vs permittivity 
% ------------------------------------------------
\subsubsection{Loss \vs permittivity}
\begin{figure}[htbp]
\begin{center}	
\includegraphics[width = 0.8\textwidth]{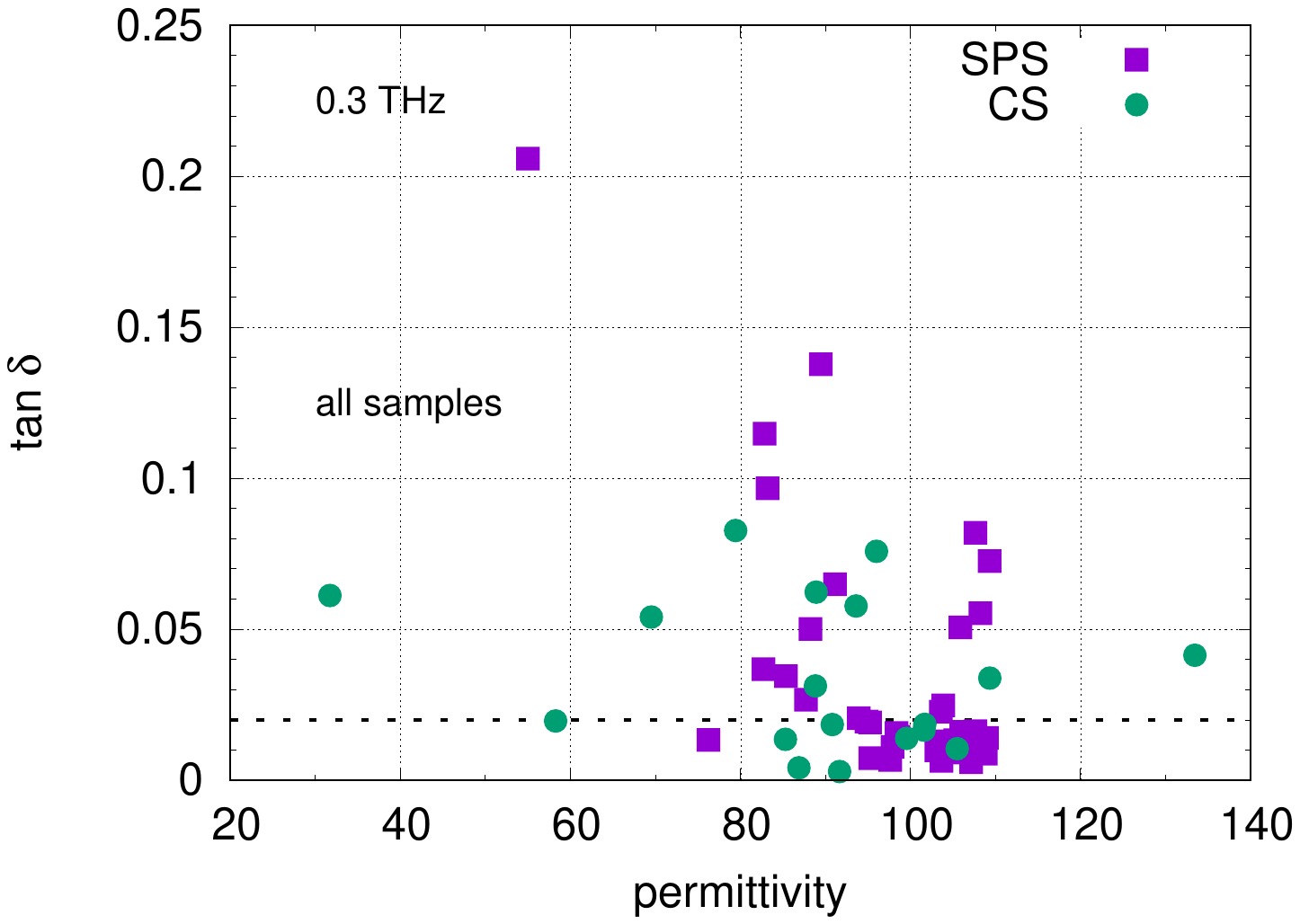}
\caption{Loss tangent ($\tandel=\epsilon''/\epsilon'$) in function of the permittivity $\epsp$ of all the samples at 0.3\thz: SPS samples in violet and CS samples in green. The horizontal dotted line indicates the threshold of loss tangent $\tandel = 0.02$.}
\label{fig:permittivity}
\end{center}
\end{figure}

The loss tangent \vs the permittivity from the\thz-TDS measurements at 0.3\thz is reported in \fig{fig:permittivity}, and it can be also noticed that both the processes allow to fabricate samples with low loss ($\tandel < 0.02$) and high permittivity ($\epsp \simeq 100$). Nonetheless, most of SPS samples satisfy the two criteria:  $\varepsilon' \simeq 100$ and $\tandel \leq 0.02$, and, notably, the SPS samples exhibit the lowest loss associated to the highest permittivity. 
Two CS samples exhibit a lower loss tangent ($\tandel = 3\dix{-3}$ and $\tandel = 4\dix{-3}$, however, their respective permittivity are $\epsp=91$ and $\epsp=87$.
The similar figures at 0.6\thz and at 1\thz are reported in the supplementary material %\;\seefsm{S-fig:tgdelta600} 
\seefsmd{S-fig:tgdelta600}{S-fig:tgdelta1000} and it shows that the criteria of the loss tangent threshold is more difficult to satisfy at 0.6\thz because the loss tangent induced by the first optical phonon $\phonon$ is higher. And that it is not achieved at 1\thz. 

% -------------------------------------------------------
% --- Dielectric Function
% -------------------------------------------------------
\subsubsection{Dielectric Function}

The \thz spectrum measurements were then compared with a theoretical model, namely the pseudo-harmonic (PH) model, which is convenient since our measurements were carried out in the lower end of the\thz band (<1.5\thz). {Actually, considering that we have characterised our samples in the lower THz band, the influence of higher frequencies resonances due to the LO$_1$ phonon or higher order phonon modes have a very low impact on the dielectric response of \tiod below 1.5\thz. So we only consider the effect of the first transverse optical phonon TO$_1$ mode.} We built this model based on our measurements taken from two single crystals rutile \tiod samples (Surfacenet GmbH), oriented in the $<001>$ and $<100>$ directions. The theoretical dielectric function in each direction is given by 
\begin{equation}\label{eq:ph}
\varepsilon^*_{\parallel, \perp}(\omega) = \epsp(\omega) + \imath\epspp(\omega)
= \varepsilon_\infty + \frac{ \omega^2_{\phonon}\!\cdot\left(\varepsilon_0 -  \varepsilon_\infty\right)} {\omega^2_{\phonon} - \omega^2 - \imath \gamphonon\omega}.
\end{equation}
It involves four parameters. Along the rapid axis, we used the values of $\varepsilon_0$ and $\varepsilon_{\infty}$ from Matsumoto \etal\cite{jjap47_matsumoto} and Traylor \etal, \cite{prb3_traylor}, while for the slow axis, we adapted those from Kanehara \etal \cite{jcsj123_kanehara}, to determine $\epsperp$ and $\epspar$, respectively. These are reported in the supplementary material \seetabsm{S-tab:ph_parameters}. The figure of the two corresponding dielectric functions is also reported in the supplementary material \seefsm{S-fig:func_diel_rutile}, and it shows good agreement between the model and the measurements along both directions {($\parallel$, $\perp$)} of the single crystal.

\begin{figure}[ht]
\centering
\begin{subfigure}{0.5\textwidth}
    \includegraphics[width=1.2\textwidth]{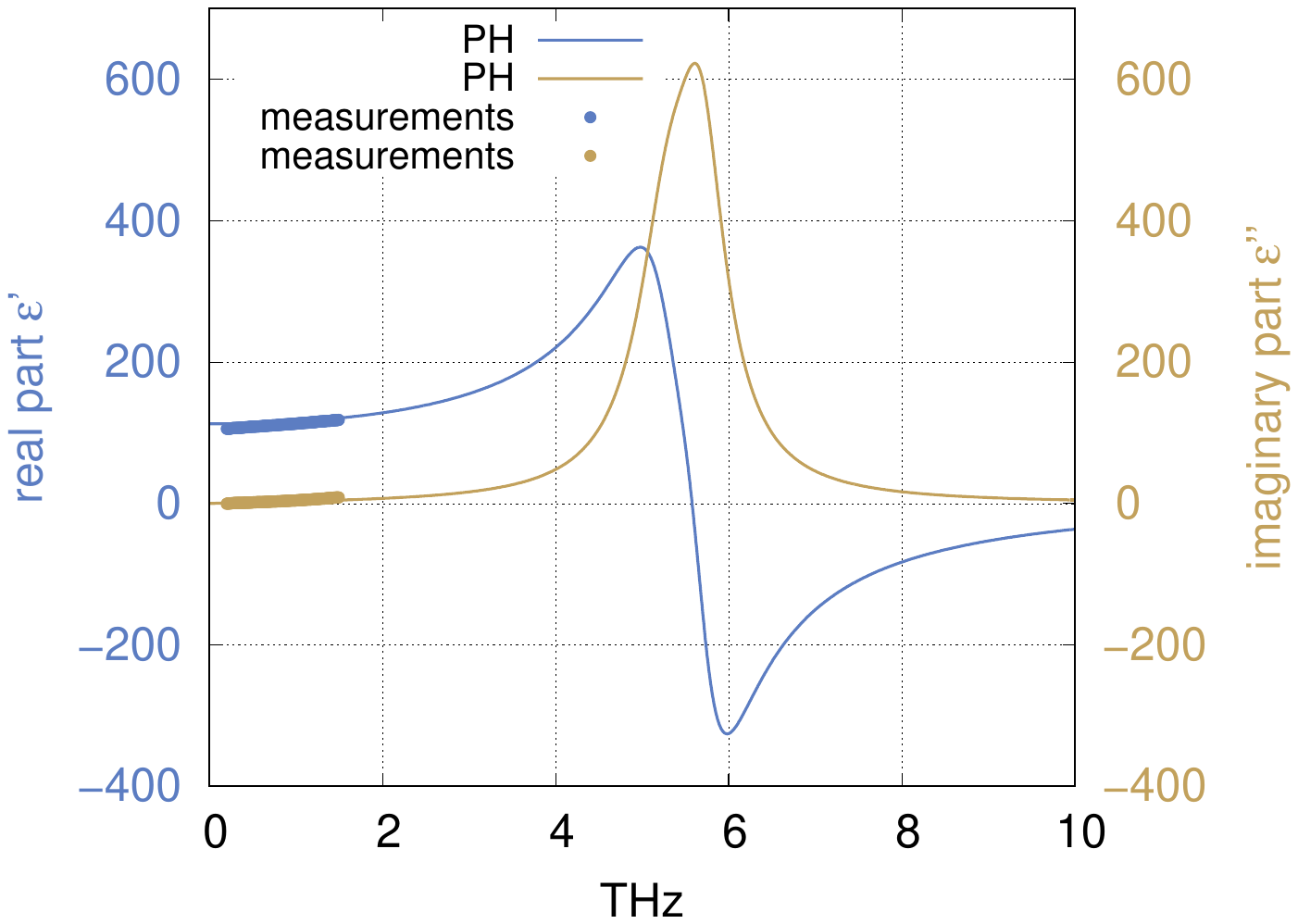}
    \caption{}
    \label{fig:fpsq1}
\end{subfigure}
\\
\begin{subfigure}{0.5\textwidth}
    \includegraphics[width=1.2\textwidth]{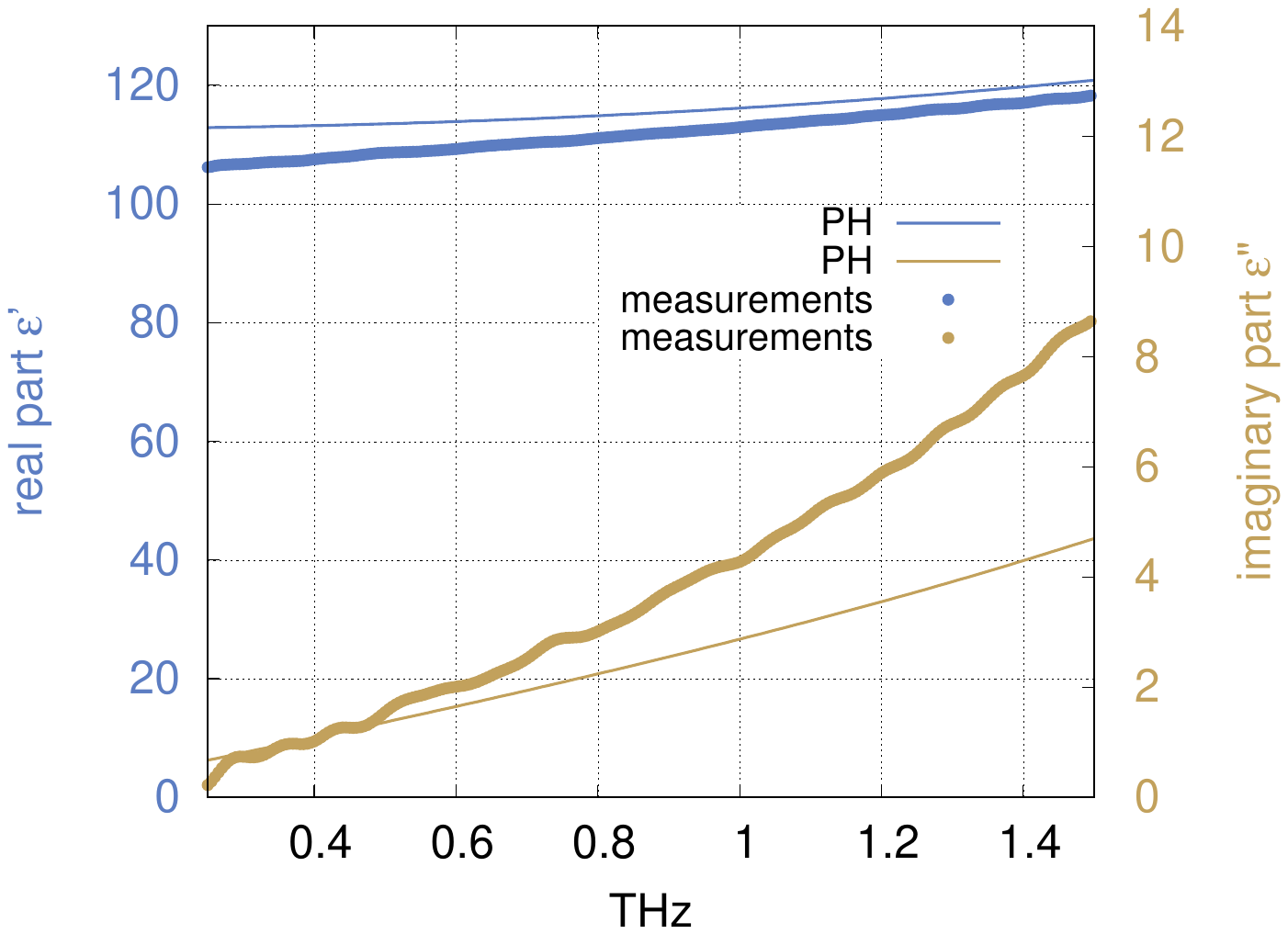}
    \caption{}
    \label{fig:fpsq2}
\end{subfigure}
\caption{Dielectric function $\varepsilon^*=  \epsp + \imath \epspp$ of \tiod sample 1.3 \seetab{tab:tandelta} sintered by SPS. Experimental spectrum (dots) in the THz range compared {with the dielectric permittivity $\epsilon_{mean}$ of a randomly oriented \tiod rutile polycrystal simulated using a Pseudo-Harmonic model (PH lines) \seeeqd{eq:ph}{eq:mean} }
%??? with the Pseudo-Harmonic model (lines) \seeeq{eq:ph}. 
(a) $[0,10]$\thz range. The maximum of the imaginary part indicates the frequency of the first optical phonon $\phonon$ at 5.7\thz.  (b) Zoomed view in the $[0.2,1.5]$\thz range. }
\label{fig:ph}
\end{figure}

Next, the dielectric function $\varepsilon* = \varepsilon' + \imath \varepsilon"$ of our polycrystalline pellets is found under the assumption that the samples consist of randomly oriented rutile grains, which leads to the mean permittivity given by\;\cite{ jjap47_matsumoto}

%given by the mean permittivity

%Since our pellets are polycrystalline, we consequently define an average permittivity \cite{ jjap47_matsumoto}. 
%
%;(see also Fig.\;2 in \cite{djemmah2022processing}
\begin{equation}\label{eq:mean}
\varepsilon_{mean} =  \frac{2}{3}\varepsilon^*_{\perp} + \frac{1}{3}\varepsilon^*_{\parallel}.% $\varepsilon^* = 
\end{equation}
%%which is the dielectric function $\varepsilon* = \varepsilon' + \imath \varepsilon"$ of the polycrystalline pellet. 
This is reported in \fig{fig:ph} and compared with the measurements, and it shows good agreement. At 0.3\thz, the measured real part $\epsp$ is 95\% of that from the PH model, which is consistent with the relative density (0.99) of the considered sample. %{However, at higher frequencies, the value of $\epspp$  for the pellet is increasing about twice faster as the reference model. We ascribe this behaviour to the presence of ionized defects, specifically to the concentration of oxygen vacancies (VO)??? and Ti interstitials (Tii) !!! 24,25; which are responsible for possible additional backscattering losses that add to the losses due to the excitation of the first transverse optical phonon.  Moreover, it hase been reported that the electrical properties of single crystals differ from those of polycrystalline specimens due to the presence of grain boundaries. And are porous.}
The imaginary part $\epspp$ of the measured dielectric function increases about twice faster as the PH model, which we ascribe to the presence of ionized defects, specifically to the concentration of oxygen vacancies (\ce{V_O}) and \ce{Ti} interstitials (\ce{Ti_i})\,\cite{rsc46_setvin, sr7_yu}. {These are responsible for possible additional backscattering losses that add to the losses due to the excitation of the first transverse optical phonon TO$_1$. Because of their different structures, the dielectric properties of single crystals differ from those of polycrystalline specimens due to the presence of grain boundaries\cite{nowotny_csr42}. Moreover, the latter are porous.}

%(V$_O$) and Ti interstitials (Ti$_i$)\,\cite{rsc46_setvin}
% Four Parameter Semi-Quantum model (FPSQ)\;\cite{jjap47_matsumoto}

% -------------------------------------------------------
% --- Permittivity \vs density
% -------------------------------------------------------

\subsubsection{Effect of the sintering and annealing temperatures on the loss of SPS samples}
From the qualitative results of \fig{fig:permittivity}, we further investigated the effect of the SPS sintering temperature and that of the annealing on the loss tangent of the samples, the permittivity being a less critical parameter\cite{sr7_yu}. We considered twenty-seven samples realized from three different commercial powders, whose characteristics are reported in \tab{tab:powders}. The samples were considered in three sets from the SPS sintering temperature: \ts=\cels{1100}, \cels{1150}and \cels{1200}. Each set comprises three subsets of samples made of the same powders, i.e., nine samples in each set. In each subset, two samples were annealed: one at \tann=\cels{1000} and the other at \cels{1100}. 
% samples made of the same powder were fabricated, among which two samples have been annealed at two different temperatures: \tann=\cels{1000} or \cels{1100}, one being thus not annealed. 
This annealing was carried out during 60 hours under air, because the re-oxidation kinetics of \tiod remain very low in this range of temperatures.
All these SPS samples were next characterized at 0.3\thz and the results are reported in \tab{tab:tandelta}. Several points can be noticed: 
\begin{enumerate}[label=(\roman*) ]
\item {Without annealing, the powder whose specific surface area \sbet is the lowest exhibits the lowest loss tangent $\tandel$. Indeed, it is less affected by the reduction phenomena during the sintering process. Its defect concentration \titp is lower. 
}
%is the lowest is the less affectec 

%Without annealing, the powder with lower specific surface (or lower chemical reactivity) is less affected to the reduction phenomena duirng the sintering step. Then, the powder  with lower specific surface show the lower dielectric loss (or lower the Ti3+ defect concentration). 
%\item Without annealing, the purest powder leads to the lowest loss tangent. % ($\tandel\leq$ 0.022)
\item The annealing of the samples leads to a significant reduction in losses, by a factor 2 up 16, which demonstrates the importance of this post-sintering process, because it reduces the defect concentration (\ce{Ti^3+}) and the oxygen vacancies concentration (\ce{V_O}) by the re-oxidation of \tiod.

%Post-sintering annealing treatments reduce defect concentration by reoxidation of TiO2 and a decrease in oxygen vacancy concentration.
%\item \tsps leads to ????
\item The lowest SPS temperature (\tsps = \cels{1100}) followed by an annealing leads to the loss tangent below the threshold $\tandel \leq 0.02$ (black and blue bold fonts in \tab{tab:tandelta}). 
%\item 
\item The process \tsps = \cels{1100} \& \tann= \cels{1000} leads to the lowest loss tangent ($\tandel \leq 0.016$) (blue bold font in \tab{tab:tandelta}). Notice that one sample exhibits a loss tangent as low as $\tandel = 6.10^{-3}$ at 0.3\thz and $\tandel = 18.10^{-3}$ at 0.6\thz\cite{sr7_yu}. Its particule size \dbet is the lowest. These SPS conditions, at lower temperature, limit the creation of \titp defects or low reduction of \tiod.

%{Une explication possible ? ???} .%\item only the pair of values \tsps=\cels{1100} and \tann=\cels{1000} leads to the loss tangent lower than threshold 0.02 for three samples made by the same process. ???
\end{enumerate}

{The Brouwer diagram, which represents the conductivity $\sigma$ in fonction of the oxygen partial pressure $\ppod$ at various temperatures, gives an insight of the role of the sintering and annealing temperatures\;\cite{pssb244_nowotny, jpcb110_nowotny, jms23_balachandran}. It shows that, at a given partial oxygen pressure $\ppod$, sintering and annealing at lower temperatures lead to lower conductivity (lower defect concentration \titp). Consequently, sintering at \cels{1100} followed by annealing at \cels{1000} is the optimal process considering our three sets of samples. The Brouwer diagram with corresponding partial oxygen pressures of SPS sintering and annealing is shown in the supplementary material \seefsm{S-fig:brouwer}.
}

\begin{table}[ht]
    \centering
    \begin{tabular}{@{}llll@{}}
     \toprule
     Producer (reference) &  Purity (\%) & \sbet (m$^2$/g) & \dbet ($\mu$m)\\
     \midrule
Sigma Aldrich (10897) & 99.995 & 0.16 & 19\\
Sigma Aldrich (42681) & 99.8 & 4 & 4.4\\
HP2 & 99 & 6.3 & 1\\
       \bottomrule
    \end{tabular}
    
    \caption{Characteristics of the \tiod powders used to fabricate the SPS samples. The values of the specific surface area (\sbet) are those given by the producer. Using the Brunauer-Emmett-Teller' method (BET), the particle size \dbet was determined\;\cite{pst32_courard}.%{(Est-ce que c'est clair ? Pr\'eciser ce qu'est \dbet  ? ???)}
    }
\label{tab:powders}
\end{table}

% ------------------------------------------------
% ----  Permittivity and loss tangent in function of the sintering temperatures (annealed samples) 
% ------------------------------------------------
% ------------------------------------------------
% --- table 2
% ------------------------------------------------
\begin{table}[ht]
  \centering
  \ra{1.3}
   \hspace*{-3cm}
%\begin{minipage}{\textwidth}
\noindent\makebox[\textwidth]{ 
\resizebox{1.1\textwidth}{!}{
 \begin{tabular}{lccclccclccc}
    \toprule
    \multicolumn{4}{c}{set 1: SPS at \cels{1100}} & \multicolumn{4}{c}{set 2: SPS at \cels{1150}} & \multicolumn{4}{c}{set 3: SPS at \cels{1200}} \\
%    \multicolumn{4}{c}{SPS at \cels{1100}} & \multicolumn{4}{c}{SPS at \cels{1150}} & \multicolumn{4}{c}{SPS at \cels{1200}} \\
    \cmidrule(rl){1-4 } \cmidrule(rl){5-8 } \cmidrule(l){9-12 }
&  \multicolumn{3}{c}{annealing} & & \multicolumn{3}{c}{annealing} & & \multicolumn{3}{c}{annealing} \\
   \cmidrule(rl){2-4 } \cmidrule(rl){6-8 } \cmidrule(l){10-12 }
%
%ref & none & \cels{1000} & \cels{1000} & &\cels{1000} & & & & & \\
 subset & none & \cels{1000} & \cels{1100} & subset & none & \cels{1000} & \cels{1100} & subset & none & \cels{1000} & \cels{1100} \\
    \cmidrule(rl){1-4 } \cmidrule(rl){5-8 } \cmidrule(l){9-12 }
 1.1 (10897) & 0.022&\txcb{\textbf{0.009}}&\textbf{0.019} & 2.1 (10897) & 0.016&0.009&0.015& 3.1 (10897) & 0.024&0.034&0.016 \\
 1.2 (42681) & 0.082&\txcb{\textbf{0.016}}&\textbf{0.019} & 2.2 (42681) & 0.072&0.026&0.050& 3.2 (42681) & 0.055 & 0.013 & 0.016 \\
 1.3 (HP2) & 0.096&\txcb{\textbf{0.006}}&\textbf{0.013} & 2.3 (HP2) & 0.050&0.013&0.014 & 3.3 (HP2) & 0.064&0.020& 0.027\\
\bottomrule
    
   \end{tabular}
   }
   }
% \end{minipage} 
    \hspace*{-3cm}
  \caption{\label{tab:tandelta}Loss tangent ($\tandel= \epsilon''/\epsilon'$) at 0.3\;THz in fonction of the annealing temperature of twenty seven samples, which were considered in three sets from the three different SPS temperatures: \ts=\cels{1100}, \cels{1150} and \cels{1200}. Each set comprises three subsets of samples fabricated from the same powder (see\;\tab{tab:powders}). In each subset, two samples were annealed: one at \tann=\cels{1000} and the other at \cels{1100}. 
%They were fabricated from three different powders (see\;\tab{tab:powders}) and sintered by SPS at three different temperatures (\ts=\cels{1100}, \cels{1150} and \cels{1200}), and further annealed at \tann=\cels{1000}, \cels{1100}, or not.  
The bold font indicates that the loss tangent is below the threshold $\tandel=0.02$ at a given annealing temperature for the three samples fabricated by the same process (sintering then annealing). The process \tsps=\cels{1100} \& \tann= \cels{1000} leads to loss tangent below 0.016 (blue column).}
\end{table}

% ------------------------------------------------
% ---- Effective parameters and negative index 
% ------------------------------------------------
% ---------------------------------------------------------------------------
\subsection{Effective parameters and negative index}
\begin{figure}[ht]
\centering
\begin{subfigure}{0.45\textwidth}
    \includegraphics[width=1.2\textwidth]{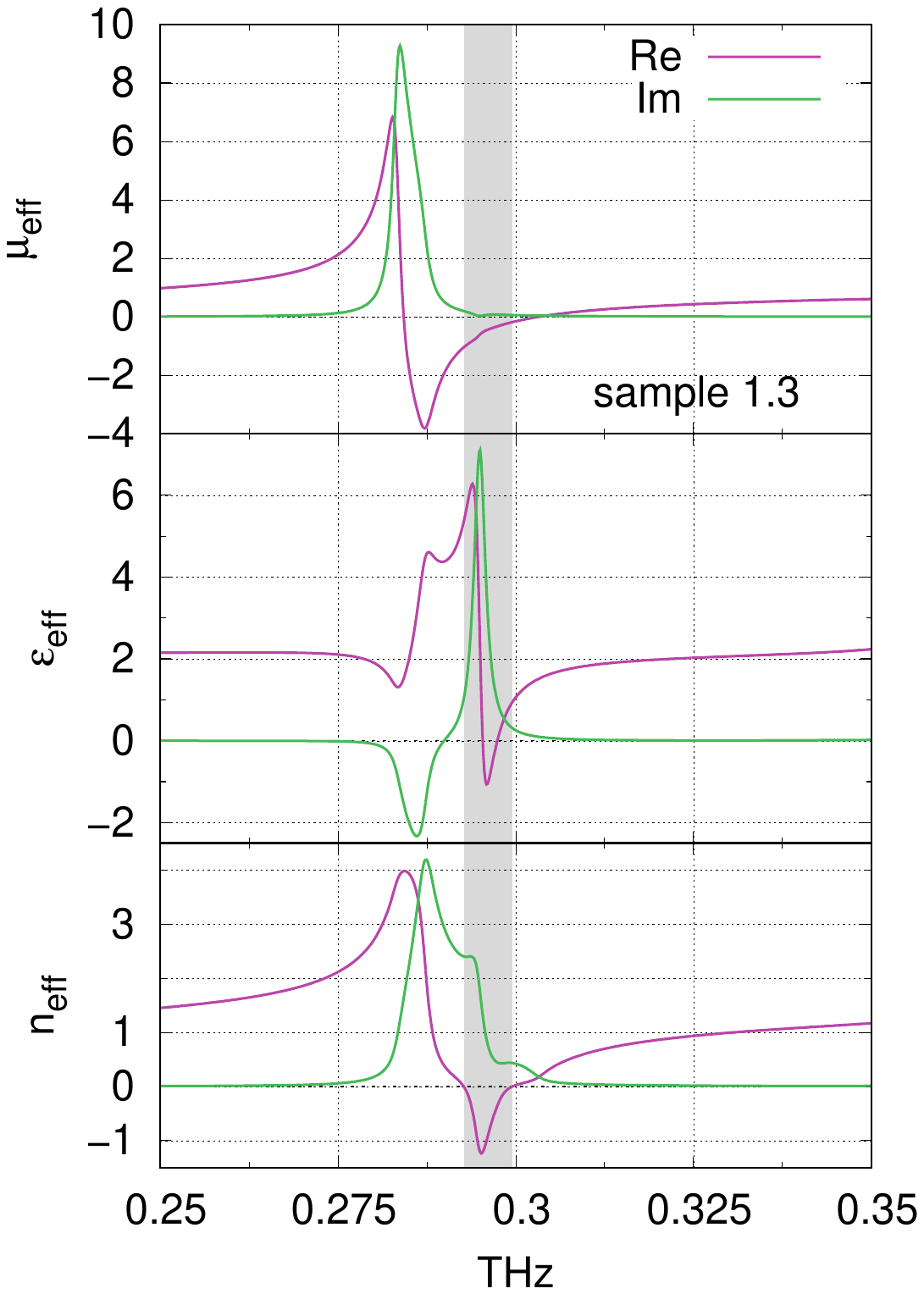}
    \caption{}
    \label{fig:param_eff_300}
\end{subfigure}
\hfill
\begin{subfigure}{0.45\textwidth}
    \includegraphics[width=1.2\textwidth]{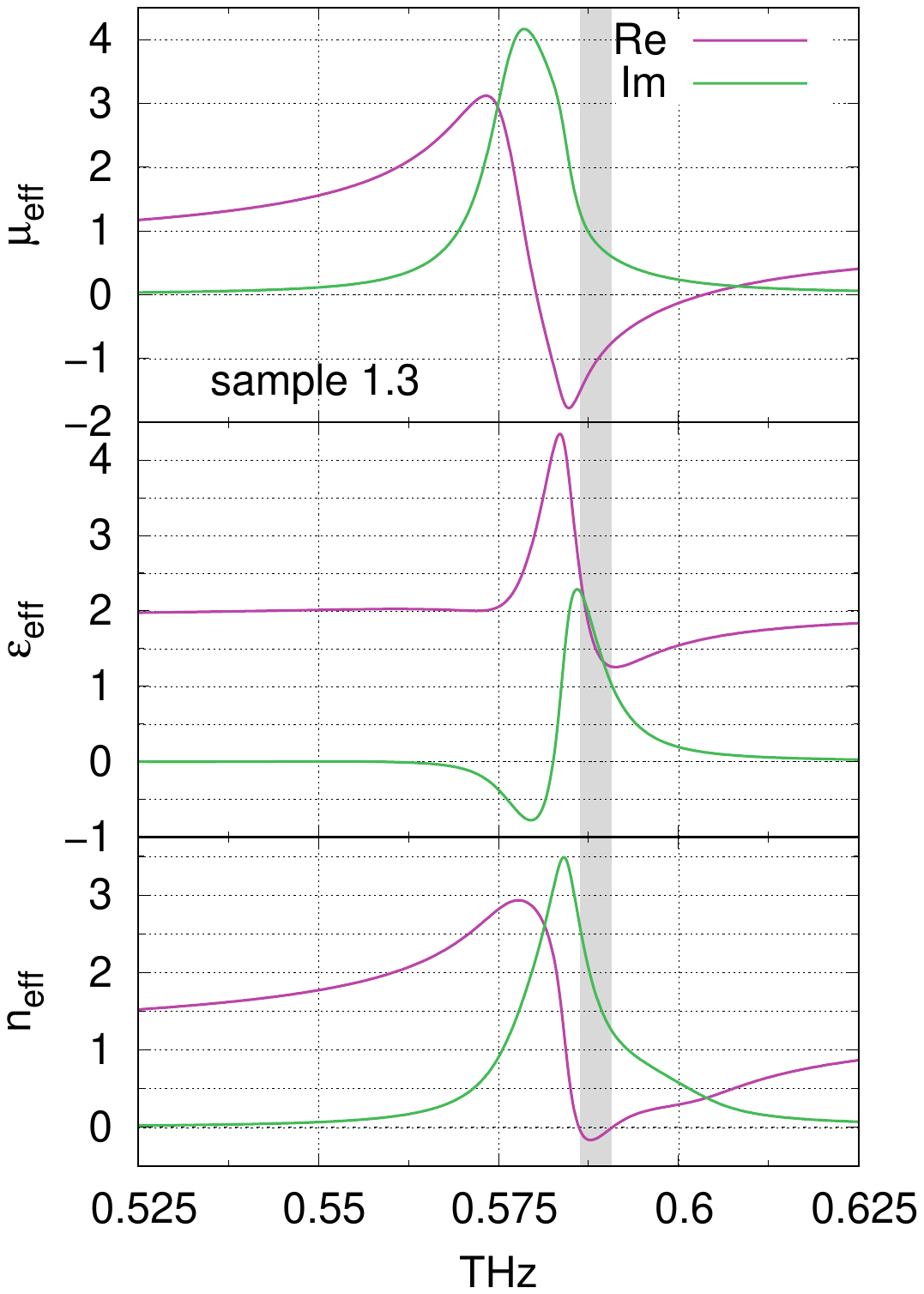}
    \caption{}
    \label{fig:param_eff_600}
\end{subfigure}
\caption{Effective parameters (real and imaginary parts): permeability $\mueff$, permittivity $\epseff$ and index $\neff$ from sample 1.3 \seetab{tab:tandelta}, extracted from numerical simulations: (a) The two modes of Mie resonances are around  0.3\thz, the minimum of the effective index is $\neffmin= -1.23$ at 0.295\thz; (b) The two modes of Mie resonances are around at 0.6\thz, the minimum of the effective index is $\neffmin= -0.16$ at 0.587\thz. The shaded areas mark the frequency range in which the effective index $\neff$ is negative.}
\label{fig:param_eff}
\end{figure}

To design the device, we used the experimental EM characteristics of the sample of subset 1.3, that was sintered by SPS at \cels{1100} and then annealed at \cels{1000}under air during 60 hours (blue column in \tab{tab:tandelta}). These are $\epsp = 103, \tandel = 0.006$ at 0.3\thz and  $\epsp = 105, \tandel = 0.018$ at 0.6\thz, respectively\;\seef{fig:ph}.
The calculated effective parameters: permeability $\mueff$, permittivity $\epseff$ and index $\neff$ in function of the frequency are reported in \fig{fig:param_eff}. 
The maximum of the imaginary part of the effective permeability $\mueff$ indicates the frequency of resonance of the magnetic resonator (magnetic mode \tuu), while the maximum of the imaginary part of the effective permittivity $\epseff$ indicates that of the electric resonator (electric mode \tud). 
These are \fuu = 0.283\thz and \fud = 0.294\thz, respectively. %These are equal to that of Cohn's, because the distance $p$ between two resonators is large enough to avoid coupling between modes 
The gap ($\Deltafs = 11$\ghz \vs $\Deltafc = 13$\ghz) with that calculated from Cohn's model (argon) is the evidence of the mode coupling\;\cite{epjam5_marcellin}. At 0.6\thz, these frequencies are \fuu = 0.586\thz and \fud = 0.594\thz, respectively. The gap ($\Deltafs = 8$\ghz \vs $\Deltafc = 12$\ghz) is greater, since the periodicity $p= 120$\mum is  lower in that case, reinforcing the coupling between the two modes\;\cite{epjam5_marcellin}. 
Besides, metamaterials exhibit negative effective index of refraction as Depine and Lakhtakia's Condition (DLC) is satisfied\;\cite{motl41_depine}, that is, when
% ---Depine and Lakhtakia's Condition
\begin{equation}\label{eq:dlc}
\epseff'\,\mueff'' + \epseff''\,\mueff' < 0,
\end{equation} 
% ---
where the prime and the double prime denote the real and imaginary parts, respectively. {It takes loss into account.} %These simulations show that negative index is achievable. 
The corresponding frequency ranges are shown by the shaded areas of \fig{fig:param_eff}. Its minimum values are $\neffmin = -1.23$ at 0.295\thz and $\neffmin = -0.16$ at 0.587\thz. The corresponding bandwidths are 6.8\ghz at 0.3\thz and 4.5\ghz at 0.6\thz, respectively. These simulations thus demonstrate that negative index is achievable at THz from  the \tiod pellets we realized by SPS followed by an annealing at both frequencies.  They next require to be micro-structured at the tens of micrometers scale with a micrometer resolution\;\cite{ijmwt_amina}. {We are currently implementing micro-molding of ceramic, using polymer molds, which is a microscale manufacturing process\;\cite{ijmwt_amina}. This process has the advantage of facilitating the precise realization of components by accurately reproducing the specific characteristics of a polymer mold.} In the supplementary material (see~figures\:\ref{S-fig:param_eff_sample11} and \ref{S-fig:param_eff_sample12} in the supplementary material), we also report the same simulations of the two other samples (samples 1.1 and 1.2) fabricated by the same processes (blue column in \tab{tab:tandelta}): these show that if negative index can be achieved at 0.3\thz, it is not at 0.6\thz, since the loss tangent $\tandel$ is larger than the loss tangent threshold 0.02 at this latter frequency, because of the first optical phonon $\phonon$. Nevertheless,  it can be noticed that sample 1.1 exhibits a near-zero index at 0.6\thz, whereas sample 1.3, whose loss $\tandel$ is higher, does not \seefsm{S-fig:sample_at1_600}.

% ------------------------------------------------
% ---- Effective parameters:
% ------------------------------------------------

% -------------------------------------------------------
% --- Conclusion
% -------------------------------------------------------
\section{Conclusion}
To develop All-Dielectric Metamaterials operating at THz frequencies, we have studied the EM properties of \tiod pellets we fabricated from commercial powders. 
From our previous results, these devices require a high permittivity ($\epsilon'\simeq 100$) and low loss ($\tandel \leq 0.02$). 
We first compared two sintering processes: Conventional Sintering (CS) and Spark Plasma Sintering (SPS), in function of the temperature. Then, we investigated the effect of a post-sintering annealing on the loss in function of the temperature. The pellets were characterized in the THz frequency range by THz Time Domain Spectroscopy (THz-TDS). We show that the two criteria are achievable by SPS and annealing. The process \tsps = \cels{1100} and \tann= \cels{1000} yields the best results. Notably, a sample exhibiting a loss tangent $\tandel = 6.10^{-3}$ at 0.3\thz and $\tandel = 18.10^{-3}$ at 0.6\thz, with respective permittivities $\epsp=103$ and $\epsp=105$ has been realized. By adjusting the coupling between the modes, near-zero index should be allowed. From these measurements, we designed All-Dielectric Metamaterials and show that they exhibit negative index in the THz range. Realizing All-Dielectric Metamaterials next necessitates to micro-structure the bulk material at the tens of micrometers scale with a micrometer resolution.

\section{Acknowledgements}
D.A.D. thanks the Mission pour les Initiatives Transverses et Interdisciplinaires (MITI) of the CNRS for the scholarship. The XRD characterizations have been carried out thanks to the french RENATECH network.

\section{Funding}
This work has been financed by the french Agence Innovation D\'efense (via the ANR Dispont project) and the MITI (CNRS). 

\section{Author contributions statement}

D.A.D. designed and simulated the devices, D.G. and P.-M. G. fabricated the samples, F.-B. prepared the samples, D.A.D. and J.-F.R. characterized the samples,  E.A. supervised the study. All authors have analysed the results, contributed to the manuscript and reviewed it.

%Must include all authors, identified by initials, for example:
%A.A. conceived the experiment(s), A.A. and B.A. conducted the experiment(s), C.A. and D.A. analysed the results. All authors reviewed the manuscript.

\section{Additional information}
\textbf{Competing interests}: The authors declare none.

The corresponding author is responsible for submitting a \href{http://www.nature.com/srep/policies/index.html#competing}{competing interests statement} on behalf of all authors of the paper. This statement must be included in the submitted article file.

\section{Data availability}
Data underlying the results presented in this paper are not publicly available at this time but may be obtained from the authors upon reasonable request.

\section{Supplemental document}
See Supplemental document for supporting content.

\bibliography{jecs_amina}
\end{document}